\documentclass{camera}
\usepackage{graphicx}  

\begin{document}

%
\title{Present status of coupled-channels calculations 
for heavy-ion subbarrier fusion reactions}

%
\author{K. Hagino$^{1,2,3}$ and J.M. Yao$^{1,4}$}

%
\organization{$^1$ Department of Physics, Tohoku University, 
Sendai 980-8578, Japan \\
$^2$ Research Center for Electron Photon Science, Tohoku University,\\
Sendai 982-0826, Japan \\
$^3$ National Astronomical Observatory of Japan, 2-21-1 Osawa, Mitaka,
Tokyo 181-8588, Japan \\
$^4$ School of Physical Science and Technology,
Southwest University, Chongqing 400715, China}
\maketitle

\begin{abstract}
The coupled-channels method has been a standard tool in analyzing 
heavy-ion fusion reactions at energies around the Coulomb barrier. 
We investigate three simplifications usually adopted in the coupled-channels 
calculations. These are i) the exclusion of non-collective excitations, 
ii) the assumption of coordinate independent coupling strengths, 
and iii) the harmonic oscillator approximation for multi-phonon excitations. 
In connection to the last point, we propose a novel microscopic method 
based on the beyond-mean-field approach in order 
to take into account 
the anharmonic effects of collective vibrations. 
\end{abstract}

\section{Introduction} 

The field of heavy-ion subbarrier fusion reactions started in the late '70s, 
when a large enhancement of fusion cross sections 
was experimentally discovered 
with respect to the 
prediction of a simple potential model \cite{HT12}. 
Even though the potential model works well for light systems, such 
as $^{14}$N+$^{12}$C, it has turned out that 
it largely underestimates fusion cross sections for 
heavier systems, such as $^{16}$O+$^{154}$Sm, at energies below the 
Coulomb barrier. 
It has been well recognized by now that this large enhancement of 
subbarrier fusion cross sections is caused by the couplings of the 
relative motion between the colliding nuclei to several nuclear intrinsic 
degrees of freedom, such as low-lying collective excitations in the 
colliding nuclei as well as several nucleon 
transfer processes \cite{HT12,DHRS98,BT98,Back14}. 

Naturally a standard framework for heavy-ion subbarrier fusion has thus been 
the coupled-channels method \cite{HRK99} 
by including relevant degrees of freedom. 
This method has not only successfully accounted for the subbarrier enhancement 
of fusion cross sections for many systems but 
has also provided a natural interpretation of the so called fusion 
barrier distributions \cite{HT12,DHRS98}. 

In the coupled-channels approach for heavy-ion fusion reactions, 
the following simplifications are usually employed. 
Firstly, the model space is restricted only to low-lying collective 
excitations, excluding non-collective excitations and giant resonances. 
Secondly, the coupling strength and the excitation energy for each 
state are taken to be the same as those in an isolated nucleus 
and are assumed to be unaltered during the whole process of fusion 
reaction. Thirdly, a simple harmonic oscillator or a rigid rotor is 
assumed when multiple excitations to higher collective states are involved. 
In this contribution, we shall investigate the validity of each of these 
assumptions. 

\section{Role of non-collective excitations}

Let us start with the first assumption, that is, the 
role of 
non-collective excitations, which are usually not included 
in coupled-channels calculations. 

Low-lying collective motions are strongly 
coupled to the ground state, and also have a strong mass number 
and atomic number dependences. 
They play a major role in heavy-ion subbarrier fusion reactions, 
and are explicitly taken into account in coupled-channels calculations. 
In addition to the low-lying collective excitations, there are many other 
modes of excitations in atomic nuclei. 
Among them, 
non-collective excitations 
couple only weakly to the ground state and usually they do not 
affect in a significant way 
heavy-ion fusion reactions, even though 
the number of non-collective states is large \cite{YHR13}.  
Couplings to giant resonances 
are relatively strong due to their collective character. 
However, since their excitation energies 
are relatively large and also are smooth functions 
of mass number, 
their effects can be effectively incorporated in 
a choice of internuclear potential through the adiabatic potential 
normalization \cite{HT12}. 

\begin{figure}
\centering
\includegraphics[scale=0.5,clip]{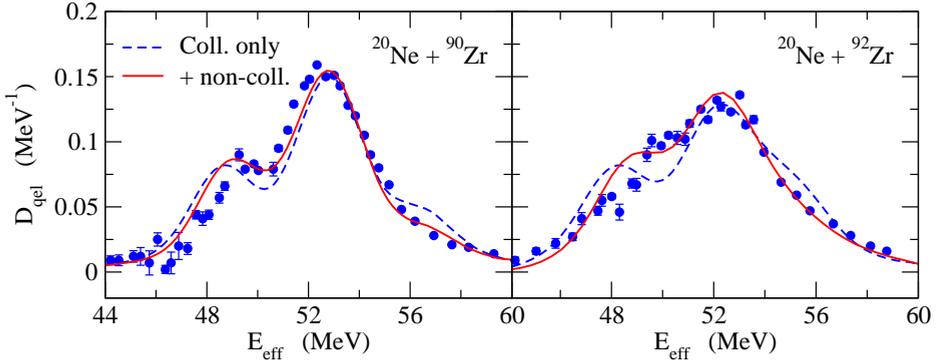}
\caption{The quasi-elastic barrier distributions for the 
$^{20}$Ne+$^{90}$Zr (the left panel) and the 
$^{20}$Ne+$^{92}$Zr (the right panel) systems. 
Here, the quasi-elastic barrier distribution is defined as 
$D_{\rm qel}=-d[d\sigma_{\rm qel}/d\sigma_{\rm R}]/dE$, 
where $\sigma_{\rm qel}$ and $\sigma_{\rm R}$ are the 
quasi-elastic and the Rutherford cross sections, respectively. 
These barrier distributions are evaluated at the scattering angle of 
$\theta_{\rm lab}$ = 150$^\circ$ and are 
plotted as a function of effective energy 
defined by 
$E_{\rm eff}= 2E\sin(\theta_{\rm c.m.}/2)/(1+\sin(\theta_{\rm c.m.}/2))$.
The dashed lines show the results of the coupled-channels calculations 
with the collective excitations in the projectile and the target nuclei, 
while the solid lines take in addition the non-collective excitations in the 
target nuclei into account with a random matrix model. 
The experimental data are taken from Ref. \cite{Piasecki}. 
}
\end{figure}

Although in most of cases, the non-collective excitations do not 
play a significant role unless the system is not very heavy, 
there are certain cases in which they play a role. 
An example is 
the $^{20}$Ne + $^{92}$Zr system, for which the quasi-elastic 
scattering at backward angles has been measured 
experimentally \cite{Piasecki}. 
Here, the quasi-elastic 
scattering refers to the sum of elastic, inelastic, and transfer 
cross sections, and is a counter part of fusion reactions \cite{HR04}. 
Figure 1 shows a comparison between the 
quasi-elastic barrier distribution, defined as 
$D_{\rm qel}=-d[d\sigma_{\rm qel}/d\sigma_{\rm R}]/dE$, where $\sigma_{\rm qel}$ and $\sigma_{\rm R}$ are the 
quasi-elastic and the Rutherford cross sections, respectively,  for the 
$^{20}$Ne + $^{90}$Zr system and that for 
the $^{20}$Ne + $^{92}$Zr systems \cite{YHR13-2}. 
One striking feature is that the experimental 
quasi-elastic barrier distribution 
for the $^{20}$Ne + $^{92}$Zr system is much more smeared 
than that 
for the $^{20}$Ne + $^{90}$Zr system \cite{Piasecki}. 
The dashed lines in the figure 
show the results of the coupled-channels calculations 
that include the rotational excitations in $^{20}$Ne as well 
as the collective phonon excitations in $^{90,92}$Zr. 
This calculation reproduces the experimental data for the 
$^{20}$Ne + $^{90}$Zr system but not for the 
$^{20}$Ne + $^{92}$Zr system.  
The solid lines, on the other hand, take into account 
also the non-collective excitations in $^{90,92}$Zr with a random matrix 
model \cite{YHR13-2}. 
One can see that the smearing of quasi-elastic barrier distribution 
for the $^{20}$Ne + $^{92}$Zr system is now well reproduced by 
the non-collective excitations of the $^{92}$Zr nucleus, whose level 
density is much larger than that of $^{90}$Zr due to the two extra neutrons 
outside the $N=50$ shell closure. 

\section{Coordinate dependent coupling strength and deep subbarrier fusion 
hindrance}

The coupled-channels approach expands the total wave function 
with the basis of isolated nuclei. 
Important inputs for coupled-channels calculations are, together with 
an internuclear potential, the excitation energy and the coupling strength 
for each excitations. 
Usually the experimental data are available for the excitation energy, 
and the coupling strength can be estimated from a measured 
electric transition probability \cite{HT12}. 
These values 
are usually employed in coupled-channels calculations 
assuming that they are not altered during the reaction process. 
This assumption has been examined recently by Ichikawa 
and Matsuyanagi \cite{IM13}. 
They have carried out random-phase approximation (RPA) calculations 
with a two-center shell model potential for 
the $^{16}$O+$^{16}$O, $^{40}$Ca+$^{40}$Ca, and 
$^{16}$O+$^{208}$Pb systems, and have demonstrated that 
the coupling strengths are indeed constant at large distances but 
they decrease appreciably in the vicinity of the touching point. 

This implies that the assumption of the constant coupling strength 
is reasonable for fusion reactions at energies around 
the Coulomb barrier. However, at deep subbarrier 
energies, the inner turning point is close to or even inner the touching 
point \cite{IHI07}, and it is important to take into account 
the effect of variation of the coupling strength. 
Notice that these are the energies at which the deep subbarrier fusion 
hindrance phenomenon has been observed \cite{Back14}. 
In fact, the finding of Ichikawa and Matsuyanagi 
provides a microscopic justification for the damping 
factor introduced phenomenologically in the adiabatic model 
for the deep subbarrier fusion 
hindrance phenomenon \cite{IHI09}. 

\section{Semi-microscopic modeling of heavy-ion fusion reactions with 
a beyond-mean field method} 

\begin{figure}
\centering
\includegraphics[scale=0.5,clip]{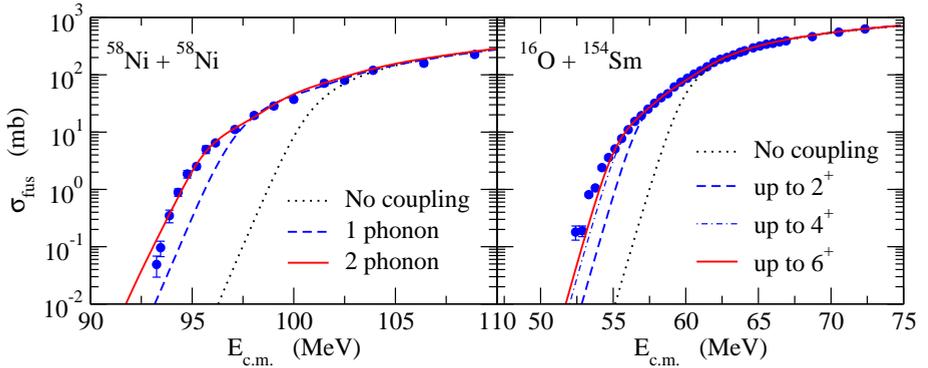}
\caption{The results of coupled-channels calculations for 
the fusion cross sections for the 
$^{58}$Ni+$^{58}$Ni (the left panel) and the 
$^{16}$O+$^{154}$Sm (the right panel) systems. 
For the $^{58}$Ni+$^{58}$Ni system, the vibrational 
coupling to multi-quadrupole-phonon 
states are considered in the harmonic oscillator approximation. 
On the other hand, for the 
$^{16}$O+$^{154}$Sm system, the rotational coupling within the 
ground state rotational 
band is taken into account. 
The experimental data are taken from Refs. \cite{Beckerman81,LDH95}. 
}
\end{figure}

In heavy-ion fusion reactions at energies around the Coulomb 
barrier, multiple excitations to higher members of collective 
states, such as multi-phonon states and high-spin states in the 
ground state rotational band, often play an important role. 
Figure 2 shows typical examples for this. The left and the right 
panels show fusion cross sections for the 
$^{58}$Ni+$^{58}$Ni and the 
$^{16}$O+$^{154}$Sm systems obtained with the coupled-channels 
calculations. For the former system, the vibrational coupling 
to quadrupole phonon states are considered, while the rotational coupling 
within the ground state rotational band is taken into account for the 
latter system. One can see that for both the systems 
the coupling to the first excited state is insufficient and 
the couplings to the higher members are necessary in order to account 
for the subbarrier enhancement of fusion cross sections. 
This feature has been demonstrated beautifully also through the 
analyses of fusion barrier distributions \cite{DHRS98,LDH95,Stefanini95}. 

In order to take into account those multiple excitations in 
coupled-channels calculations, 
one usually uses the rigid rotor model for deformed nuclei 
and the harmonic oscillator model for vibrational 
nuclei \cite{HT12}. 
In reality, however, most nuclei have neither a 
pure harmonic oscillator
spectrum nor a pure rigid body rotational band, although 
the rigid rotor approximation is reasonable for medium-heavy and heavy 
deformed nuclei. 
For example, the $^{58}$Ni nucleus, which has usually been
considered to be a typical vibrational nucleus,
does not exhibit a level spectrum characteristic to the
harmonic vibration, {\it e.g.}, the degeneracy of the two-phonon
triplet is considerably broken.
Moreover, a recent theoretical calculation 
also indicates that the $B(E2)$ strengths
among the collective levels in $^{58}$Ni 
deviate largely from the harmonic oscillator limit \cite{Yao15}. 

There are several theoretical ways to describe an anharmonic vibration. 
Among them, 
a multi-reference density-functional
theory (MR-DFT) 
has been rapidly developed 
for the past decade \cite{Bender03,Yao14,Vretenar15}.
This method is based on the so called beyond-mean-field 
approximation, which incorporates 
the angular momentum and particle number projections 
as well as 
the quantum fluctuation of the mean-field wave function described by 
the generator coordinate method (GCM).

\begin{figure}
\centering
\includegraphics[scale=0.7,clip]{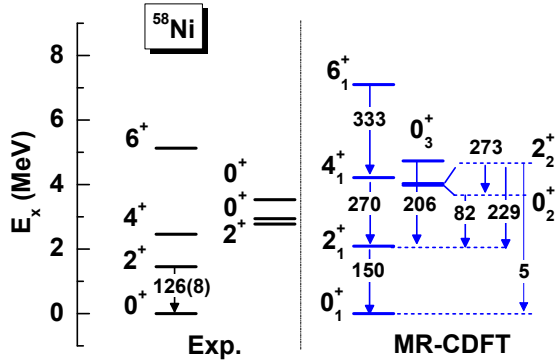}
\caption{
The low-lying energy spectra of $^{58}$Ni 
obtained with the multi-reference covariant density functional 
theory (MR-CDFT) method with the PC-PK1 force. 
The arrows indicate the 
$E2$ transition strengths, given in units of $e^2$fm$^4$. 
The experimental data
are taken from Refs. \cite{NNDC,allmond14}. } 
\end{figure}

Figure 3 shows 
the result of the MR-DFT calculation for the $^{58}$Ni nucleus. 
The calculation employs the covariant density functional theory (CDFT) with 
PC-PK1 interaction \cite{PC-PK1}, and we thus call it MR-CDFT. 
One can see that 
the main feature of the energy spectrum, as well as the $E2$ transition
strength from $2^+_1$ to $0^+_1$, are reproduced rather well.
It is interesting to notice that 
the overall pattern of $B(E2)$ values is quite different
from what would be expected for a harmonic vibrator, even though 
the excitation energies of the 4$^+_1$, 2$^+_2$, and 0$^+_2$ states
are about twice the energy of the $2^+_1$ state. 
In particular,
the $E2$ transition from the $0^+_2$ to the $2^+_1$ states 
is much smaller than that
from the $4^+_1$ and the $2^+_2$ states to the $2^+_1$ state.  
Instead, the $0^+_2$ state has a strong transition from the 
$2^+_2$ state, which clearly indicates that the $0^+_2$ state 
is not a member of the two-phonon triplet. 
Compared to the $0^+_2$ state, the $E2$
transition strength from the $0^+_3$ to the $2^+_1$ states is 
much larger and is comparable to that from the $4^+_1$ and the $2^+_2$ 
states to the $2^+_1$ state. 
This fact makes 
the $0^+_3$ state a better candidate for a member of the two-phonon 
triplets, even 
though the excitation energy is a little bit large. 

Notice that, in the harmonic oscillator limit, the $B(E2)$ value from any of 
the two-phonon triplet states to the 2$^+_1$ state is exactly twice the $B(E2)$ 
value from the 2$^+_1$ state to the ground state. 
The calculated $B(E2)$ values shown in Fig. 3, 
together with the strong transition from the 2$^+_2$ to 
the 0$^+_2$ states, 
indicate the presence of large anharmonicity in the quadrupole vibrations 
in $^{58}$Ni. 
That is, the calculated $B(E2)$ values 
are significantly quenched from the values in the harmonic limit. 

\begin{figure}
\centering
\includegraphics[scale=0.45,clip]{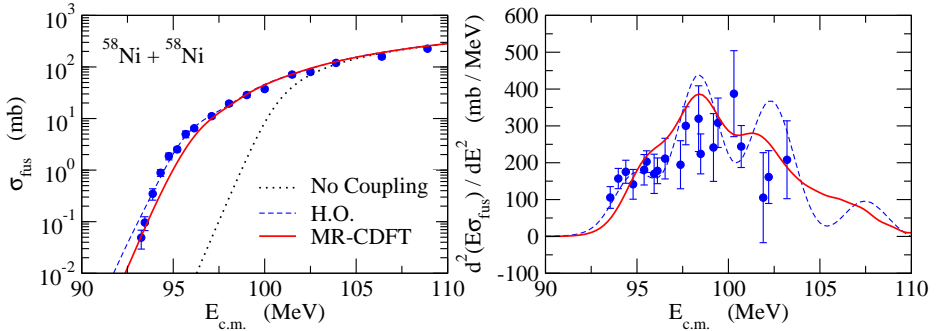}
\caption{
The fusion cross sections (the left panel) and 
the fusion barrier distributions (the right panel) 
for the $^{58}$Ni+$^{58}$Ni system. 
The dashed line is the result of the coupled-channels calculations 
including the double quadrupole phonon excitations in each $^{58}$Ni 
nucleus in the harmonic oscillator limit, while the solid line is 
obtained with the multi-reference covariant density functional 
theory (MR-CDFT) method by including 
the 0$^+_1$, 2$^+_1$, 0$^+_3$, 2$^+_2$, and 4$^+_1$ states. 
The dotted line in the left panel denotes 
the result in the absence of the channel couplings. 
The experimental data are taken from Ref. \cite{Beckerman81} for 
the fusion cross sections and from Ref. \cite{Stefanini95} 
for the fusion barrier distribution. 
}
\end{figure}

We can now ask ourselves 
how 
the deviation of the spectrum from the harmonic limit 
affects the subbarrier fusion reactions 
of Ni isotopes. 
Figures 4 shows the fusion 
cross section $\sigma_{\rm fus}(E)$ 
and the fusion barrier distribution 
$D_{\rm fus}(E) = d^2(E\sigma_{\rm fus})/dE^2$ \cite{DHRS98,RSS91}
for the $^{58}$Ni+$^{58}$Ni reaction. 
The dashed line shows the result of the 
coupled-channels calculations including up to the double phonon states  
in the harmonic oscillator limit. 
All the mutual excitations between the projectile and the target nuclei
are included. 
On the other hand, the solid line in the figure is obtained 
with the coupling strengths calculated with the 
MR-CDFT method while adopting a phenomenological Woods-Saxon potential. 
To this end, we include 
the 0$^+_1$, 2$^+_1$, 0$^+_3$, 2$^+_2$, and 4$^+_1$ states in the 
coupled-channels calculations.
Again, all the mutual excitation channels are taken into account. 
See Ref. \cite{HY15} for the details of the calculations. 
For a comparison, the figure also shows the result of no-coupling limit by 
the dotted line.
One can see that the calculations in the harmonic limit overestimate
fusion cross sections at the two lowest energies, while the MR-CDFT
calculations underpredict fusion cross sections around 95 MeV.
For the energy dependence of fusion cross
sections, shown in terms of fusion barrier distribution 
in the right panel of the figure, the MR-CDFT calculation leads to a minor
improvement by considerably smearing each peak.

We have carried out similar calculations also for 
the $^{40}$Ca + $^{58}$Ni system \cite{HY15}. 
We have observed again also for this system 
that the anharmonicity effect in $^{58}$Ni smears
the fusion barrier distribution, leading to a better agreement with the
experimental fusion barrier distribution
as compared to the results in the harmonic
oscillator limit. 

\section{Summary}

The coupled-channels approach has been a standard tool in the field 
of heavy-ion subbarrier fusion reactions. 
We have discussed three assumptions used in usual coupled-channels 
calculations. 

We have first discussed the role of non-collective 
excitations. Usually, coupled-channels calculations take into 
account a few selected low-lying collective states, neglecting the couplings 
to non-collective states. We have demonstrated that even though the 
non-collective 
excitations can indeed be neglected in many cases, there are certain systems 
which show those effects. One example is 
$^{20}$Ne + $^{92}$Zr system, which shows a considerably smeared barrier 
distribution as compared to the $^{20}$Ne + $^{90}$Zr system. 
By explicitly including many non-collective excitations, we have demonstrated 
that they indeed smear the barrier distribution for 
$^{20}$Ne + $^{92}$Zr system while the effect is much smaller for 
the $^{20}$Ne + $^{90}$Zr system. 

We have next discussed 
two important input quantities in 
coupled-channels calculations, that is, the coupling strength and the 
excitation energy for each state. 
In most of coupled-channels calculations, these quantities are assumed to be 
unchanged during the reaction 
process and are taken to be constants. 
The recent RPA calculations have indicated that this is the case at 
large distances but they decreases considerably around the touching 
point. We have argued that this leads to important consequences for 
the deep subbarrier hindrance of fusion cross sections. 

We have then proposed a semi-microscopic approach to heavy-ion subbarrier 
fusion reactions. 
The basic idea of this approach is to combine a multi-reference 
density functional theory (MR-DFT) to a coupled-channels calculation. 
The MR-DFT provides transition strengths 
among collective states without resorting to the harmonic oscillator model 
or the rigid rotor model. 
The advantages of this approach include i) deviations from the harmonic 
limit as well as the rigid rotor limit can be taken into account, 
ii) it can therefore be applied also to transitional nuclei, which 
show neither the vibrational nor the rotational characters, 
and iii) a natural truncation is introduced in the coupling schemes. 
We have applied this approach to 
the $^{58}$Ni+$^{58}$Ni and $^{40}$Ca+$^{58}$Ni fusion 
reactions, and have 
found that the anharmonicities smear the fusion barrier distributions, 
somewhat improving the agreement with the experimental data. 

One of the important current issues in nuclear reaction theory is 
to develop a microscopic framework starting from the nucleon degree of 
freedom. It has however been extremely challenging to construct a fully 
microscopic theory which is applicable to heavy-ion subbarrier fusion 
reactions, thus to many-particle quantum tunneling. 
We believe that the semi-microscopic approach presented in this paper 
provides an important step towards this direction. 

\section*{Acknowledgements}
We thank N. Rowley, T. Ichikawa, and S. Yusa 
for useful discussions. 
This work was partially supported by the National Natural 
Science Foundation of China under 
Grant Nos. 11305134,  11105111, and the Fundamental Research 
Funds for the Central 
University (XDJK2013C028).


\begin{thebibliography}{99}

\bibitem{HT12} Hagino K. and Takigawa N., Prog. Theor. Phys.
{\bf 128} (2012) 1001. 

\bibitem{DHRS98}Dasgupta M., Hinde D.J., Rowley N. and
Stefanini A.M., Annu. Rev. Nucl. Part. Sci. {\bf 48} (1998) 401. 

\bibitem{BT98}Balantekin A.B. and Takigawa N.,
Rev. Mod. Phys. {\bf 70} (1998) 77. 

\bibitem{Back14} Back B.B., Esbensen H., Jiang C.L. and Rehm K.E.,
Rev. Mod. Phys. {\bf 86} (2014) 317. 

\bibitem{HRK99}Hagino K., Rowley N. and Kruppa A.T., 
Comp. Phys. Comm. {\bf 123} (1999) 143. 

\bibitem{YHR13}
Yusa S., Hagino K. and Rowley N., 
Phys. Rev. C {\bf 88} (2013) 044620. 

\bibitem{Piasecki}
Piasecki E. {\it et al.}, Phys. Rev. C{\bf 80} (2009) 054613. 

\bibitem{HR04}
Hagino K. and Rowley N., Phys. Rev. C{\bf 69} (2004) 054610. 

\bibitem{YHR13-2}
Yusa S., Hagino K. and Rowley N., 
Phys. Rev. C {\bf 88} (2013) 054621. 

\bibitem{IM13}Ichikawa T. and Matsuyanagi K., Phys. Rev. C{\bf 88} 
(2013) 011602 (R); arXiv: 1506.07963. 

\bibitem{IHI07} Ichikawa T., Hagino K. and Iwamoto A.,
Phys. Rev. C{\bf 75} (2007) 057603; 
Phys. Rev. C{\bf 75} (2007) 064612.

\bibitem{IHI09} 
Ichikawa T., Hagino K. and Iwamoto A.,
Phys. Rev. Lett. {\bf 103} (2009) 202701. 

\bibitem{Beckerman81}
Beckerman M. {\it et al.}, 
Phys. Rev. C{\bf 23} (1981) 1581. 

\bibitem{LDH95}Leigh J.R. {\it et al.}, 
Phys. Rev. C{\bf 52} (1995) 3151.

\bibitem{Stefanini95} Stefanini A.M. {\it et al.}, 
Phys. Rev. Lett. {\bf 74} (1995) 864. 

\bibitem{Yao15} Yao J.M., Bender M. and Heenen P.-H.,
Phys. Rev. C{\bf 91} (2015) 024301. 

\bibitem{Bender03}
Bender M., Heenen P.-H. and Reinhard P.-G.,
Rev. Mod. Phys. {\bf 75} (2003) 121.

\bibitem{Yao14}
Yao J.M., Hagino K., Li Z.P., Meng J. and Ring P.,
Phys. Rev. C{\bf 89} (2014) 054306.

\bibitem{Vretenar15}
Vretenar D., the contribution in this conference. 

\bibitem{HY15}
Hagno K. and Yao J.M., 
Phys. Rev. C{\bf 91} (2015) 064606. 

\bibitem{NNDC} National Nuclear Data Center, http://www.nndc.bnl.gov/.

\bibitem{allmond14}Allmond J.M. {\it et al.}, 
Phys. Rev. C{\bf 90} (2014) 034309.

\bibitem{PC-PK1} Zhao P.W., Li Z.P., Yao J.M. and Meng J.,
Phys. Rev. C{\bf 82} (2010) 054319.

\bibitem{RSS91}Rowley N., Satchler G.R. and Stelson P.H., 
Phys. Lett. {\bf B254} (1991) 25. 


\end{thebibliography}
\end{document}